# Early-Time Flux Measurements of SN 2014J Obtained with Small Robotic Telescopes: Extending the AAVSO Light Curve

Björn Poppe[1], Thorsten Plaggenborg[1], WeiKang Zheng[2], Isaac Shivvers[2], Koichi Itagaki[3], Alexei V. Filippenko[2], and Jutta Kunz[1]

[1] University Observatory, Carl von Ossietzky University Oldenburg, Germany

[2] Department of Astronomy, University of California, Berkeley, CA 94720-3411, USA

[3] Itagaki Astronomical Observatory, Teppo-cho, Yamagata 990-2492, Japan

**Summary**

In this work, early-time photometry of supernova (SN) 2014J is presented, extending the AAVSO CCD database to prediscovery dates. The applicability of NASA's small robotic MicroObservatory Network telescopes for photometric measurements is evaluated. Prediscovery and postdiscovery photometry of SN 2014J is measured from images taken by two different telescopes of the network, and is compared to measurements from the Katzman Automatic Imaging Telescope and the Itagaki Observatory. In the early light-curve phase (which exhibits stable spectral behavior with constant color indices), these data agree with reasonably high accuracy (better than 0.05 mag around maximum brightness, and 0.15 mag at earlier times). Owing to the changing spectral energy distribution of the SN and the different spectral characteristics of the systems used, differences increase after maximum light. We augment light curves of SN 2014J downloaded from the American Association of Variable Star Observers (AAVSO) online database with these data, and consider the complete brightness evolution of this important Type Ia SN. Furthermore, the first detection presented here (Jan. 15.427, 2014) appears to be one of the earliest observations of SN 2014J yet published, taken less than a day after the SN exploded.

**Introduction**

Supernova (SN) 2014J in the nearby galaxy M82 is widely believed to be one of the most important Type Ia SN explosions of the last few decades, and appears to be the nearest known SN Ia to explode since SN 1972E or, perhaps, Kepler's SN (Foley *et al.* 2014). The object was discovered by S. J. Fossey and his students on January 21.805, 2014 (UT dates are used throughout this paper). A number of prediscovery observations were subsequently reported (Fossey *et al.* 2014; Ma *et al.* 2014; Denisenko *et al.* 2014), and the first-light time was estimated by Zheng *et al.* (2014) to be January 14.75 ± 0.21.

Despite the fact that M82 was regularly monitored every few days, several robotic SN search programs failed to detect SN 2014J automatically (probably owing to the complex host-galaxy morphology; see, for instance, Zheng *et al.* 2014). Other early-time observations, including spectroscopy and photometry, have been published by several groups (e.g., Goobar *et al.* 2014; Marion *et al.* 2014; Foley *et al.* 2014; Tsvetkov *et al.* 2014). While these studies indicate that SN 2014J was a normal SN Ia, it also exhibited remarkably severe visual extinction and reddening. According to Foley *et al.* (2014), the observations may be explained by a combination of extinction in the M82 interstellar medium and scattering processes in circumstellar material.

We published prediscovery observations of SN 2014J collected with the NASA MicroObservatory Network telescopes (MOs) in the online American Association of Variable

Star Observers (AAVSO) database (unfiltered light curve saved under the observer name "UniOl1"). The MOs monitor M82 as part of the "interesting objects" catalogue with a cadence of 1 day, if weather permits. The MOs consist of a network of robotic telescopes developed by the Harvard-Smithsonian Center for Astrophysics (Sadler *et al.* 2001) to enable world-wide access to astronomical images for students and interested amateurs over the World Wide Web. The telescopes are located and maintained at professional observatories in the United States. They can be controlled remotely, or specific objects can be chosen from a set of predefined suggestions covering examples of the most important astrophysical objects, such as galaxies, star clusters, and nebulae. The predefined suggestions are imaged each night, sometimes even with more than one telescope (as is the case for M82), and all data are archived for at least two weeks in an online accessible database. In this work, we show that these publicly available data can be used to complement scientific campaigns such as light-curve measurements of supernovae and variable stars.

The aim of this paper is therefore twofold: (a) we show that the robotic MOs are able to perform accurate photometric measurements of variable stars, offering the potential to provide valuable data for the AAVSO database or for inclusion in larger survey campaigns; and (b) we use SN 2014J data from the Katzman Automatic Imaging Telescope (KAIT; Filippenko *et al.* 2001) and the Itagaki Observatory (Japan) along with data from the MOs to construct measurements in the Rc passband, which can be used to extend the AAVSO light curves into the prediscovery time and may also act as baseline data for further analysis and comparisons with other measurements or explosion models.

**Material and Methods**

**a) Telescopes Used in This Study**

*MicroObservatory Net (MO)*

A detailed technical description of the system is given by Sadler *et al.* (2001), and so we only describe the most important features of the MOs. The main optical system is composed of Maksutov telescopes with a primary mirror diameter of 15.24 cm (6 inches) and a nominal focal length of approximately 540 mm. The MO CCD cameras contain Kodak KAF 1400 chips, with a resolution of 1000 × 1400 pixels, where each pixel is 6.9 μm on a side. Two different fields of view are realized by using different image scales. Either a binning of 2 × 2 pixels is used (1 degree field of view), or only the inner part of the image is analyzed (1/2 degree field of view), resulting in a scale of 5.0 or 2.5 arcsec per pixel, respectively.

The telescopes are equipped with a filter wheel holding several astronomical filters. The images of M82 were taken either with an *RGB* filter set or a clear filter; we used the latter as an approximation of the unfiltered measurements of SN 2014J from other telescopes. For our study two different telescopes were used, named "Ben" and "Cecilia," both located at the Whipple Observatory, Arizona, USA. All images were taken with an exposure time of 30 s. Both telescopes take daily images of M82. Our earliest detection of SN 2014J was found in an image taken by Cecilia on January 15, about 6 days before the discovery. It is one of the earliest detections published so far.

Overall, more than 170 observations of SN 2014J by the MOs were collected and analyzed for this study, although only the early prediscovery data have been uploaded to the AAVSO database.

*Katzman Automatic Imaging Telescope (KAIT)*

Early-time photometry of SN 2014J was obtained with KAIT as part of the Lick Observatory Supernova Search (LOSS; Filippenko *et al*. 2001; Leaman *et al*. 2011). The system consists of a 0.76 m primary mirror with a focal length of approximately 6.2 m. The camera is a

MicroLine77 (Finger Lakes Instrumentation, Lima, NY, USA) with square 24 μm pixels and an image size of 500 × 500 pixels, resulting in a scale of about 0.79 arcsec per pixel.

*Itagaki Astronomical Observatory, Japan*

Additional early-time data were taken with a 0.5 m telescope by K. Itagaki (Zheng *et al*. 2014). This telescope has a focal length of approximately 5.0 m, and it uses an N-83E camera from Bitran (Saitama, Japan) with a KODAK KAF-1001E chip (1024 × 1024 pixels; 24 μm per pixel) with a scale of about 1.45 arcsec per pixel.

*Spectral Sensitivity*

As described by Zheng *et al*. (2014), the quantum-efficiency curve of the KAIT system reaches half-peak values at ~3800 and 8900 Å, and the Itagaki system has half-peak values at ~4100 and 8900 Å. The half-peak values for the MOs nominally are ~4100 and 7800 Å (Sadler *et al*. 2001). The slightly lower efficiency of the MOs at longer wavelengths results in a closer approximation to the standard Johnston *R*-band filter (half-peak values at ~5600 and 7900 Å). In their paper on the first-light time estimates, Zheng *et al*. (2014) showed that the overall agreement between the KAIT and Itagaki telescopes is better than 0.02 mag over a *B*–*R* difference of ~0.9 mag, with a systematic offset of 0.02 mag. We expect the differences between the MOs and the other telescopes to be slightly larger.

**b) Data Analysis**

*MicroObservatories for Photometric Evaluations*

For an initial verification of the MO system, catalogued stars of all available images were analyzed with Astrometrica (Version 4.1.0.293, by H. Raab Linz, Austria). Astrometric reduction and photometry were done with the help of the UCAC4 star catalogue *R*-band filter data. Astrometrica automatically determines the photometric zeropoint $M_{sys}$ though a combination of a Gaussian fit and an aperture-photometry procedure. First, to calculate the nominal positions and to identify reference stars, a two-dimensional Gaussian function with a constant offset parameter is fitted as a point-spread function (PSF) for all detected sources. Second, the flux of each star is calculated through aperture photometry, using the constant offset parameter of the Gaussian fit as an estimate of the local sky background (window size of the fit equal to the size of the used aperture). The photometric zeropoint is determined by minimizing the mean deviation of all calculated magnitudes to the catalogued reference stars. For each detected star the brightness estimate $M_{est}$ is then given by

$$M_{est} = -2.5 \log (I) + M_{sys}, \quad (1)$$

where $I$ is the background-corrected signal within a predefined circular aperture (width defined by the user; for this study $r = 3$ pixels).

As a check of the image quality, for all detected catalogue stars the measured positions and brightness estimates were compared against the calculated ones; results are presented in the next section.

*Photometry of SN 2014J*

In each image $I_{SN}$, the signal content of SN 2014J was again obtained by the method described in the previous section. Background light from the underlying host galaxy is included in the estimate for the local sky background (as above), and is automatically subtracted. Using the signal content of a photometric comparison star $I_{comp}$ and its catalogued magnitude value $M_{comp,cat}$, the magnitude of SN 2014J is then calculated from

$$M_{SN} = -2.5 \log(I_{SN}) - 2.5(\log I_{comp}) - M_{comp,cat}. \quad (2)$$

This method is equal to the one suggested by the AAVSO (Henden 2011).

A major issue in the analysis of the data is the contribution of the host-galaxy background to the photometric data. As described above, Astrometrica uses the constant offset value determined in the Gaussian-fit procedure to determine the local background. Using this procedure, uncertainties in the photometric analysis will especially become more dominant for the early phase of the SN near the detection limit. Therefore, for the earliest data, as an additional step, independent aperture-based photometry ($r = 3$ pixels) was performed to verify the evaluations done with Astrometrica. To consider the background from the host galaxy, we followed the method described by Zheng *et al*. (2014), subtracting a template of the host galaxy obtained from several observations well before the explosion (Jan. 10–13). Aperture photometry of the SN from these galaxy-subtracted images was then calibrated against the magnitude of the reference star with the same aperture-based method described above.

*Data Merging*

When the signal-to-noise ratio (SNR) is high, the dominant source of differences between two sets of stars imaged with different systems is the spectral sensitivity variations among the detectors (e.g., Riess *et al*. 1999). If data from different telescopes are to be compared or merged into one light curve, these variations in the spectral sensitivity of the systems must be taken into account, especially in the case of unfiltered measurements as used in this study. However, as long as the spectral properties of the observed object remain constant, the differences between two systems can be approximated by an additional "intersystemic" magnitude correction constant (see also Zheng *et al*. 2014). As shown below, the deviations between the intersystemic differences are comparably low and will be considered in the uncertainties of the measurements.

*Construction and Complementation of Light Curves in the AAVSO Database*

As the unfiltered passbands from the four telescopes used in this study are relatively similar to standard *R* passbands, the unfiltered data are calibrated by relative photometry to the *r'* passband of the UCAC4 catalogue. AAVSO *R*-band curves are usually given in the Cousins *Rc* passband. Owing to the small field of view of the KAIT system, no comparison stars could be found in the AAVSO database, so a direct calibration to *r'* was not possible. On the other hand, catalogued reference stars for both catalogues are available in the MO images. Therefore, observations from all 4 systems are first transformed to the *r'* passband and merged into a single light curve. In a second step, the transformation to the *Rc* passband is performed for the merged light curve of all 4 systems simultaneously by choosing the reference star from the MO images. As the light curves from the AAVSO database contain few prediscovery data, this work aims to estimate the early-time light curve of SN 2014J in the *Rc* passband, thereby complementing the existing dataset.

**Results**

*Photometric Utility of the NASA MicroObservatories*

Figure 1 shows typical images from the telescopes used for this study. As observations from two identical MOs were used, only one representative image is shown in the figure. It is clear that increasing focal lengths reveal progressively more structure in the host galaxy M82. As the scientific suitability of the Itagaki and KAIT systems have already been well established (Zheng *et al.* 2014), here we will focus on the MOs. Despite the different focal lengths of the systems, SN 2014J is clearly visible in all images.

To show the applicability of the MOs for accurate photometry, we analyzed the field stars of the images (Fig. 1, left panel) in more detail. Altogether we use 176 observations taken between January 15 and March 10 using two different telescopes named "Ben" (105 observations) and "Cecilia" (71 observations). Typically around 80 UCAC4 reference stars were detectable in each image. Astrometric reductions yielded a mean position determination better than 0.3". Instrumental zeropoint calculations were performed with Astrometrica by finding the value resulting in the smallest mean deviation between the measured and catalogued brightness values. The mean deviations for all reference stars was found to be 0.07 ± 0.04 mag for Ben and 0.07 ± 0.02 mag for Cecilia over a color index range of $B-V = 0.14$ mag through 1.18 mag. Such good agreement over this broad range shows that the unfiltered data can reasonably be compared directly to the $r'$-band values of the UCAC4 catalogue after applying an appropriate linear offset. Within the brightness range of the reference stars (8–14 mag), no instrumental deviations from linearity indicating saturation or underexposure were observed.

The field-star analysis described above can be influenced from day to day by variable stars, which may have an impact on the mean instrumental magnitude. Therefore, this global analysis of all stars acted only as a first evaluation. In the next step, brightness measurements of 4 nonvariable comparison stars from the AAVSO Variable Star Plotter database (chart number 13184CVH) in the vicinity of SN 2014J were used to evaluate the stability and accuracy of the MOs more specifically. The brightness values from the AAVSO database are given in the Cousins $Rc$ passband, whereas UCAC4 uses the Sloan $r'$ catalogue. As described above, in the first step the corresponding $r'$ values are used for relative photometry.

To detect undesired instrumental nonlinearity effects caused by underexposure or overexposure of the CCD, we include stars both brighter and fainter than SN 2014J. Stars near SN 2014J were preferentially chosen to minimize and detect effects such as sudden cloud cover or imaging problems. Observations showing deviations from the catalogued values of the reference stars larger than 0.1 mag were visually inspected for weather conditions, and if obvious cloud cover was detectable they were excluded from the study.

*Relative Photometry in the UCAC4 r' Passband*

For the KAIT and Itagaki telescopes, relative photometry in the *r'* band was performed with the same reference star as used by Zheng *et al.* (2014) (J2000 coordinates $\alpha = 09^h55^m46.1^s$, $\delta = +69°42'01.8"$, USNO-B1.0 *R2* = 15.09 mag). However, to be comparable with the MOs the instrumental magnitudes have been calibrated to the reference values of the UCAC4 catalogue in the *r'* passband (15.49 mag). As the reference star was not visible on the MO images, for both Ben and Cecilia the brightness values were calibrated against the *r'* of 000-BLG-317. As shown above, the unfiltered, *R*, and *r'* passbands are all quite similar, and assuming a simple linear offset between observations through different passbands introduces only moderate errors. Figure 2 illustrates the results of this procedure, as well as the data obtained by aperture photometry on galaxy-subtracted images in the first few days.

As Figure 2 shows, the final reduced data agree quite well, with characteristic differences less than 0.05 mag near peak brightness. In the early phase of the light curve the scatter is around 0.15 mag. Probably owing to spectral changes of the SN and the different spectral sensitivities, the agreement between the systems gets worse around the second maximum. As we seek only to obtain complementary values at early phases, we neglect observations after the first maximum.

*Analysis of the Earliest SN Detection*

Our earliest detection of SN 2014J was found in an image from Cecilia (Jan. 15.427, JD 2,456,672.928). Figure 3 shows the development of the SN after subtraction of the host galaxy. As can be seen, the images reveal a first detection on January 15 (SNR = 1.80) with

15.35 ± 0.45 mag. Though the SNR is low, and therefore the measurement uncertainty is large, the SN is clearly detected; thus, we include the data point in the light curve, and it appears to match later (high-SNR) data well. Fitting a Gaussian PSF and a linear background-galaxy light contribution, we obtain a brightness of 14.91 ± 0.2 mag; the two methods thus agree within their uncertainties.

Additionally, the values obtained by the two different photometric evaluations agree well, and they fit to the slightly later data taken by the Itagaki Observatory (see Fig. 2).

*AAVSO Database Light-Curve Evaluation*

Figure 4a shows the mean values of the data submitted by various observers to the AAVSO online database in the *RVBc* bandpasses. Averaging was performed by taking the mean values of all submitted data within a time bin of 1 day. Owing to the high reddening within the host galaxy M82, the difference between the *V* and *B* bands is quite large. From the AAVSO data the maximum was estimated by a second-order polynomial fit to be JD 2,456,690.5 ± 0.25. Zheng *et al.* (2014) estimate the first light of SN 2014J to be JD 2,456,672.25 ± 0.21; the rise time to the *B*-band maximum is therefore 18.25 ± 0.32 days. According to the AAVSO data, the uncorrected decline parameter $\Delta m_{15}(B)$ in the *B* band is approximately 1.13 ± 0.15 mag. The date for the second maximum can be estimated for the *R* band to be 25 ± 0.5 days from *B*-band maximum, or at JD 2,456,715 ± 0.5. After the second maximum, the supernova starts to fade slowly, and it declines more rapidly in *B* than in the other passbands.

As Figure 4b indicates, the color index *B−V* is around 1.25 ± 0.2 mag before or near maximum brightness, followed by the usual decline after the maximum phase.

*Transformation to the Rc Passband*

Finally, to compare and complement the AAVSO database with our own measurements, we used the tabulated difference between *r'* and *Rc* observations of the reference star 000-BLG-317 ($\Delta_{r'-Rc}$ = −0.34 mag; see Table 1) to transform all measurements to *Rc*. As described previously, because all passbands used in this project are similar, the assumption of a simple linear offset between observations through different passbands introduces only small errors. The result is shown in Figure 5. As can be seen in the overlap region, the agreement between the AAVSO data and our measurements is generally good.

**Discussion**

*MicroObservatories*

The analysis of the data herein from two MOs show that these small robotic telescopes with standard imaging parameters are well suited for astrometric and photometric measurements. Unfiltered photometry over a wide brightness range (8–14 mag) and a broad color range (*B–V* index between 0.14 and 1.18 mag) mesh well with UCAC4 data taken through the *r'* passband. Moreover, as the mean astrometric deviation is around 0.3", the accuracy requirements of the Minor Planet Center are fulfilled (better than 1"), making these devices useful for astrometry as well (IAU 2014). As another example of the scientific utility of the MOs, Fowler (2013) used MO data to determine the phase diagram of the eclipsing binary star system CQ Cep to high accuracy (better than 0.004 days, though high photometric precision was not needed for that study).

Because the MO telescopes offer only clear or *RGB* filters, their supernova science applications are restricted to first-light detections or relative photometry in the absence of strong spectral changes. As SN 2014J fulfills this latter criterion in the pre-maximum phase, we were able to complement the brightness curves of the AAVSO.

*SN 2014J*

From the AAVSO database, the *B*-band maximum was estimated to be at JD 2,456,690.5 ± 0.25. According to the data, the decline parameter is $\Delta m_{15}(B) = 1.13 \pm 0.15$ mag. The date for the second maximum can be estimated from the *R*-band data to be 25 ± 0.5 days after *B* maximum, or at JD 2,456,715 ± 0.5.

Maximum *B*-passband times were determined by Marion *et al.* (2014) to be JD 2,456,690.2 ± 0.13 (Feb. 1.74 ± 0.13) and by Tsvetkov *et al.* (2014) to be JD 2,456,691.4 (Feb. 2.9), similar to our results. The same studies derived $\Delta m_{15}$ values in the *B* band between 1.11 ± 0.02 (Marion *et al.* 2014) and 1.01 (Tsvetkov *et al.* 2014). Our value measured from the AAVSO database is consistent with the spanned interval. However, for a more detailed analysis and classification of SN 2014J (which is beyond the scope of this paper), additional corrections to $\Delta m_{15}$, accounting for extinction and host-galaxy reddening, must be included (see Phillips *et al.* 1999 or Leibundgut 2000 for an overview).

As Figure 4b indicates, the color index *B*−*V* is around 1.25 ± 0.2 mag before and near maximum brightness, increasing to approximately 2.25 around JD 2,456,670. Again, the value agrees well with the data obtained by Tsvetkov *et al.* (2014; approximately 1.3 ± 0.1 mag before maximum brightness and 2.3 mag at JD 2,456,670).

According to Leibundgut (2000), the typical absorption-corrected *B*−*V* value is −0.07 ± 0.3 mag. The remarkably high value measured for SN 2014J is undoubtedly caused by extreme dust reddening. Goobar *et al.* (2014) estimate a host-galaxy reddening for M82 of $E(B-V)_{\text{host}}$ ≈ 1.2 mag, while Schlafly & Finkbeiner (2011) report a Milky Way dust reddening of

$E(B-V)_{MW} \approx 0.14$ mag toward M82. Adoption of these values yields a corrected index of $B-V \approx -0.09$ mag, in good agreement with that of typical Type Ia supernovae. Foley *et al.* (2014) give a more detailed model based on *Hubble Space Telescope* ultraviolet spectra and multi-wavelength observations.

The color index appears to be nearly constant between first detection and *B*-band maximum, and then moves redward, thus representing the usual expected development – normal Type Ia supernovae show little $B-V$ color change until maximum light, and then exhibit a change of $-1.1$ mag over the 30 days past maximum (e.g., Ford *et al.* 1993; Leibundgut 2000).

According to Ganeshalingam *et al.* (2010), the rise times of spectroscopically normal Type Ia supernovae are generally $18.03 \pm 0.24$ days in *B*, while those exhibiting high-velocity spectral features have shorter rise times of $16.63 \pm 0.29$ days. Additionally, normal Type Ia supernovae exhibit a $\Delta m_{15}$ value in the *B* band of approximately 1.1 (Phillips *et al.* 1993). In agreement with other studies (e.g., Goobar *et al.* 2014; Tsvetkov *et al.* 2014), our derived values for $\Delta m_{15}$ ($1.13 \pm 0.15$ mag) and the rise time of 18.25 days are consistent with those of spectroscopically normal Type Ia supernovae. Although SN 2014J may be classified as a normal SN Ia, some high-velocity spectral features have been measured in its spectrum (Goobar *et al.* 2014), and some parameters are near the boundaries between the normal and high-velocity subclasses (Marion *et al.* 2014).

The MOs monitored M82 routinely each night around the same time. In this work an early-time unfiltered light curve was derived, complementing and supporting the data of Zheng *et al.* (2014).

The first image in which the SN is detected was obtained on Jan. 15.427, slightly earlier than the observation on Jan. 15.5705 presented by Zheng *et al.* (2014). The detected brightness value agrees well with their measurements and supports their estimated first-light time (Jan 14.75 ± 0.21), and it is in good agreement with the very early observations published by Goobar *et al.* (2014).

**Conclusion**

NASA's MicroObservatories can be used for photometric and astrometric observations with accuracy sufficient for some studies. Early-time photometry of SN 2014J is compared to measurements performed with the KAIT and Itagaki systems. The measurements of all systems were merged and transformed to a single photometric system so as to complement the AAVSO online database with prediscovery observations of SN 2014J. Typical parameters used to characterize Type Ia supernovae were derived from the AAVSO database, and they agree well with values from other studies.

**Acknowledgements**

The authors thank H. Raab and A. Henden for fruitful discussions. A.V.F.'s group and KAIT received financial assistance from the TABASGO Foundation, the Sylvia and Jim Katzman Foundation, the Christopher R. Redlich Fund, the Richard and Rhoda Goldman Fund, and National Science Foundation grant AST–1211916.

**References**

Denisenko, D., Gorbovskoy, E., Lipunov, V., *et al.* 2014, The Astronomer's Telegram, 5795, 1

Filippenko, A. V., Li, W. D., Treffers, R. R., & Modjaz, M. 2001, in Small-Telescope Astronomy on Global Scales, ASP Conference Series Vol. 246. Edited by Bohdan Paczyński, Wen-Ping Chen, & Claudia Lemme. San Francisco: Astronomical Society of the Pacific,

ISBN: 1-58381-084-6, p. 121

Foley, R. J., Fox, O. D., McCully, C., *et al.* 2014, MNRAS, 443, 2887

Ford, C. H., Herbst, W., Richmond, M. W., Baker, M. L., Filippenko, A. V., Treffers, R. R., Paik, Y., & Benson, P. J. 1993, AJ, 106, 111

Fossey, S. J., Cooke, B., Pollack, G., *et al.* 2014, CBET, 3792

Fowler, M. J. F. 2013, J. Br. Astron. Assoc. 123, 1

Goobar, A., Johansson, J., Amanullah, R., *et al*. 2014, ApJ, 784, L12

IAU-Minor Planet Center, Guide to Minor Body Astrometry, 2014 (http://www.minorplanetcenter.net/iau/info/Astrometry.html)

Leaman, J., Li, W., Chornock, R., & Filippenko, A. V. 2011, MNRAS, 412, 1419

Leibundgut, B. 2000, The Astronomy and Astrophysics Review, 10, 179

Ma, B., Wei, P., Shang, Z., *et al.* 2014, The Astronomer's Telegram, 5794, 1

Marion, G. H., Sand, D. J., Hsiao, E. Y., *et al.* 2015, ApJ ,798, 39

Phillips, M. M. 1993, ApJ, 413, L105

Phillips, M. M., Lira, P., Suntzeff, N. B., *et al.* 1999, AJ ,118, 1766

Riess, A. G., Filippenko, A. V., Li, W., *et al.* 1999, AJ, 118, 2675

Sadler, P., Gould, R. R., Leiker, P. S., *et al*. 2001, J. Sci. Educ. Technol., 10, 39

Schlafly, E. F., & Finkbeiner, D. P. 2011, ApJ, 737, 103

Tsvetkov, D. Yu., Metlow, V. G., Shugarov, S. Y., *et al.* 2014, Contrib. Astron. Obs. Skalnaté Pleso, 44, 1

Zheng, W., Shivvers, I., Filippenko, A. V., *et al.* 2014, ApJ, 783, L24

| Ref Star No. | Position (J2000) | $m_{UCAC4, r'}$ | Δ$m$ Ben | Δ$m$ Cecilia |
|---|---|---|---|---|
| 000-BLG-312 | 09$^h$ 55$^m$ 33$^s$ + 69° 38' 55" | 9.883 | 0.04 ± 0.05 | 0.03 ± 0.03 |
| 000-BLG-314 | 09$^h$ 54$^m$ 14$^s$ + 69° 37' 47" | 10.943 | 0.03 ± 0.05 | 0.04 ± 0.03 |
| 000-BLG-317 | 09$^h$ 56$^m$ 33$^s$ + 69° 39' 17" | 12.138 (Rc=11.798) | 0.03 ± 0.05 | 0.02 ± 0.02 |
| 000-BLG-321 | 09$^h$ 56$^m$ 38$^s$ + 68° 41' 18" | 13.314 | 0.04 ± 0.05 | 0.04 ± 0.04 |

**Table 1:** Analysis of reference stars for relative photometry of SN 2014J in the UCAC4 $r'$ passband. The value $m_{UCAC4, r'}$ is the catalogued value of the brightness, and Δ$m$ characterizes the mean deviations and standard deviations for all observations (105 for Ben, 71 for Cecilia). For 000-BLG-317 the value from the AAVSO catalogue ($Rc$) is tabulated. It is used to transform the data from $r'$ to $Rc$.

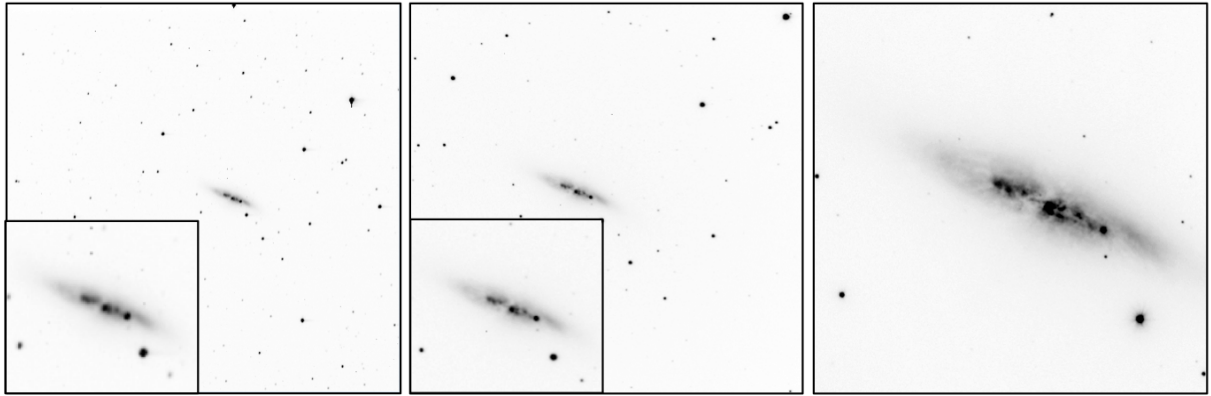

**Figure 1.** Typical images taken by the telescopes used in this study (all from Jan. 23, 2014). From left to right: An image from "Ben," representative of the MOs; an image from the Itagaki Observatory; and an image from KAIT. Inserts: enlargement of M82 for better visibility of SN 2014J.

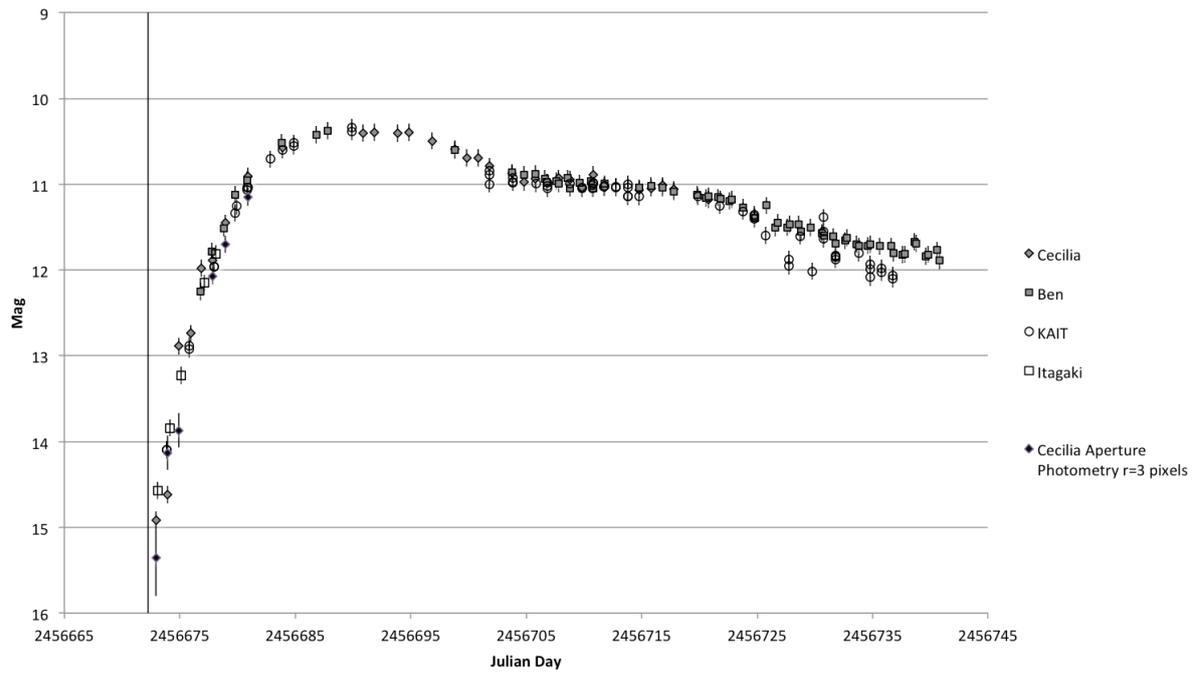

**Figure 2.** Brightness measurements of the four systems used after calibration to the *r'* passband. The vertical line indicates the first-light time estimate by Zheng *et al.* (2014).

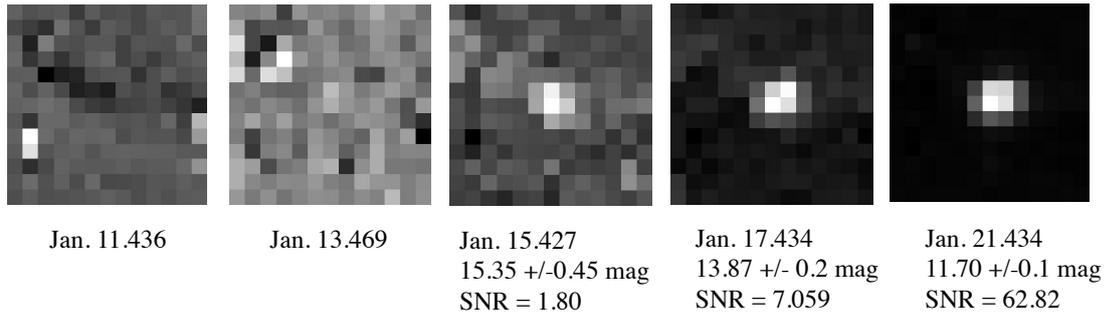

| Jan. 11.436 | Jan. 13.469 | Jan. 15.427<br>15.35 +/-0.45 mag<br>SNR = 1.80 | Jan. 17.434<br>13.87 +/- 0.2 mag<br>SNR = 7.059 | Jan. 21.434<br>11.70 +/-0.1 mag<br>SNR = 62.82 |

**Figure 3.** Development of SN 2014J in M82 after subtraction of the host galaxy (all images taken with Cecilia; the nominal detection limit at SNR = 3.0 is 14.8 mag). The first image in which SN 2014J is visible was taken on Jan. 15.427.

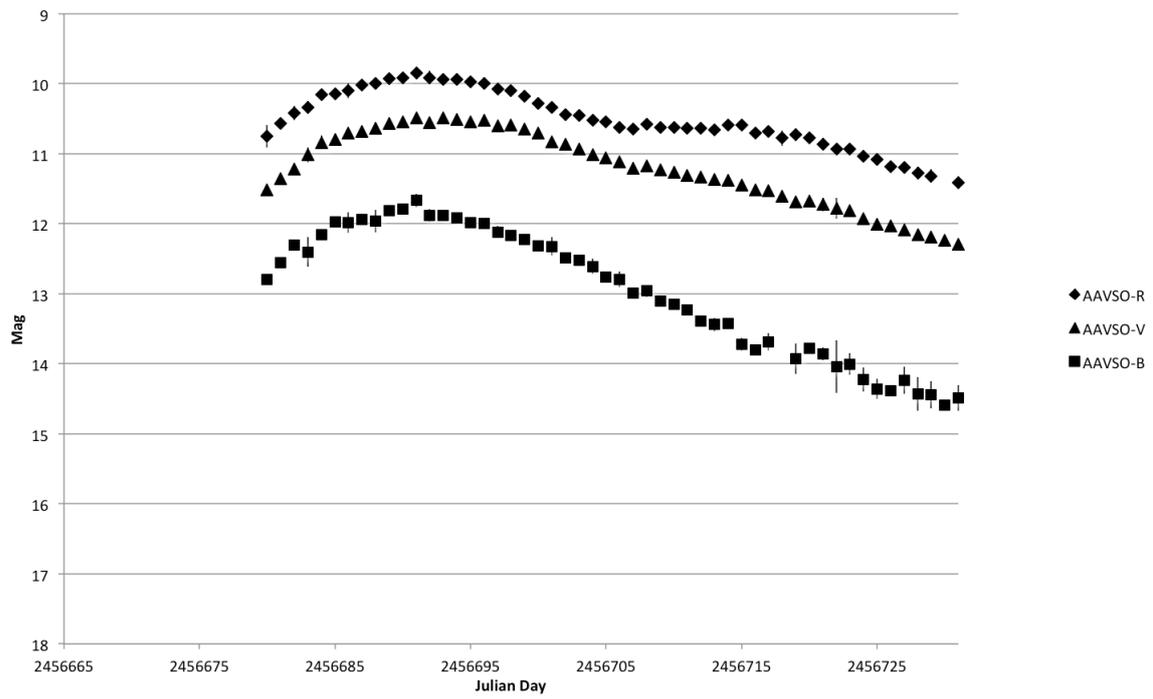

(a)

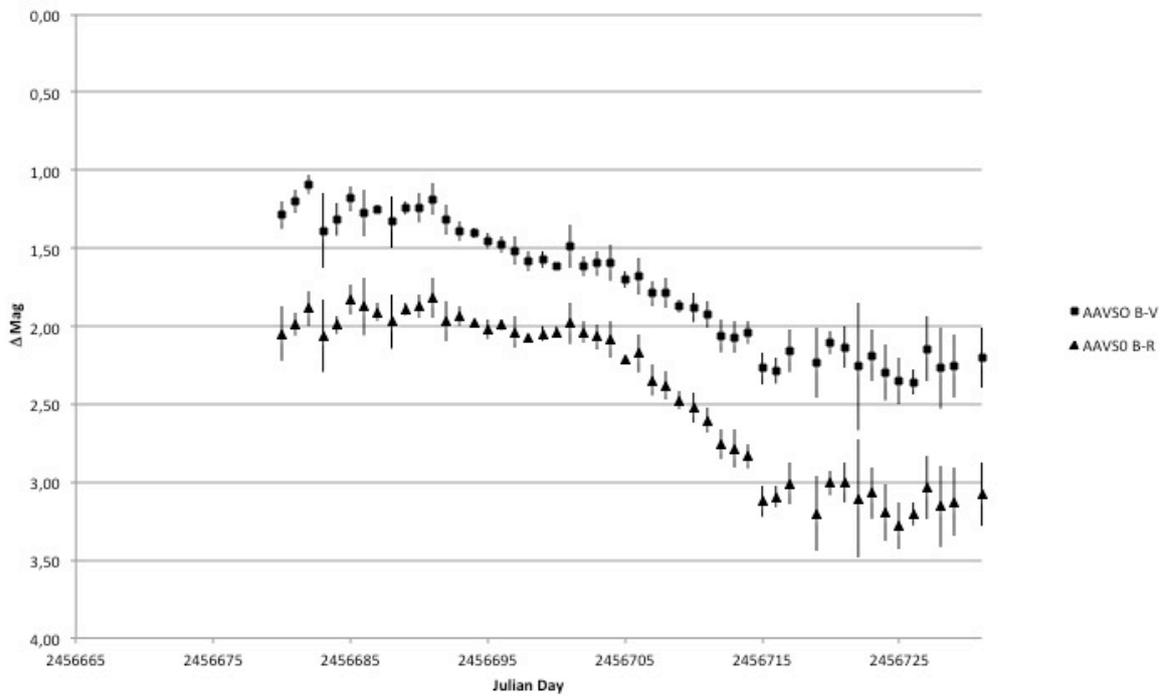

(b)

**Figure 4.** Light curve and color-index development of SN 2014J according to the AAVSO database. (a) Mean brightness and 1σ deviations in *B*, *V*, and *Rc*. (b) Mean color indices.

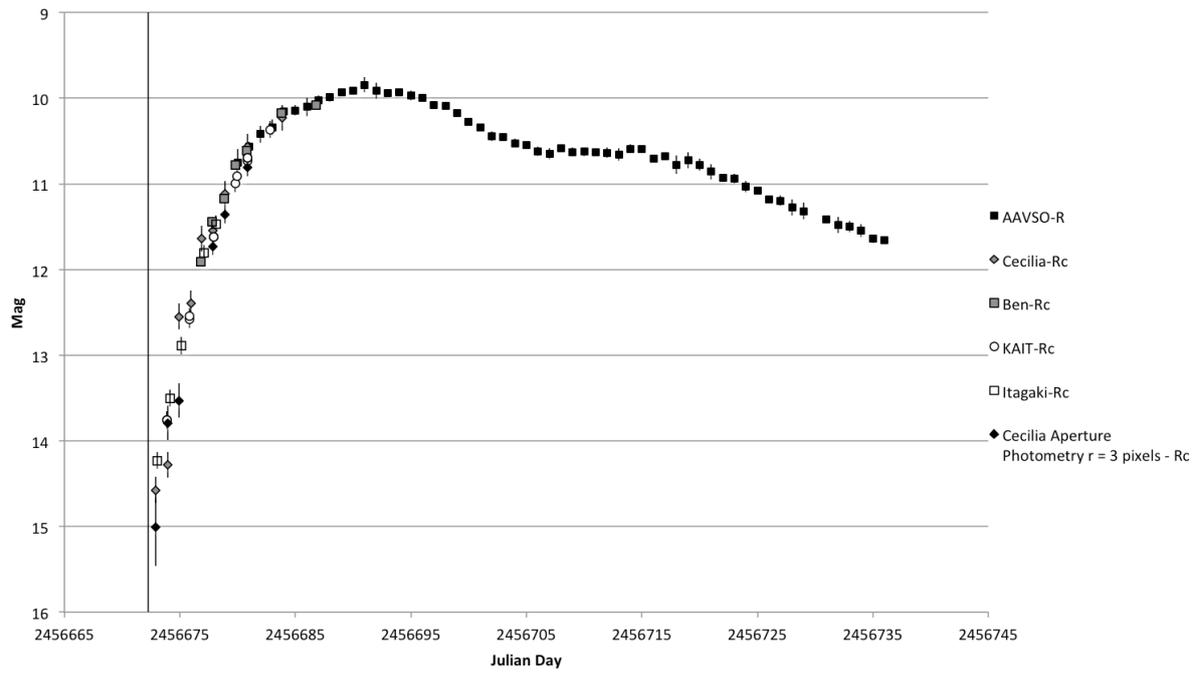

**Figure 5.** AAVSO light curve complemented by data obtained from the four telescopes used in this study.